# White Paper on DOE-HEP Accelerator Modeling Science Activities

J.-L. Vay, C. G. R. Geddes, A. Koniges – Lawrence Berkeley National Laboratory

A. Friedman, D. P. Grote – Lawrence Livermore National Laboratory

D. L. Bruhwiler – RadiaSoft LLC

J. P. Verboncoeur – Michigan State University

**Objective**

Toward the goal of maximizing the impact of computer modeling on the design of future particle accelerators and the development of new accelerator techniques & technologies, this white paper presents the rationale for: (a) strengthening and expanding programmatic activities in accelerator modeling science within the Department of Energy (DOE) Office of High Energy Physics (HEP) and (b) increasing the community-wide coordination and integration of code development.

**Expanding programmatic activities in computational accelerator science**

*Importance of modeling is on the rise -* Thanks to sustained advances in hardware and software technologies, computer modeling is playing an increasingly important role in all areas of science and technology, and this trend can be expected to continue for the foreseeable future. In application to the design of particle accelerators, this rise in importance is further fueled by the economic pressure for reducing uncertainties and the costs of development, construction and commissioning of accelerators. In light of this, it is logical to strengthen and expand programmatic activities of particle accelerator modeling within the HEP community. This increased focus on computational activities is all the more timely as computer architectures are transitioning to new technologies [1,2] that require the adaptation of existing - and emergence of new - algorithms and codes [3].

*Optimizing resources via a comprehensive and balanced development strategy* – Maximizing the overall scientific output per dollar means maximizing the usability of the pool of codes while minimizing spending on code development and support. Maximizing usability involves an aggregate maximization of effectiveness (completeness, accuracy of solution), efficiency (time to solution), and ease of use (learnability, error tolerance, versatility, visibility of functionalities, etc.). Minimizing spending on code development implies the reduction of duplication with commensurate increases in modularity and code interoperability. Optimizing overall efficiency implies a comprehensive and balanced strategy that evaluates codes not only based on performance, effectiveness, ease of use, level of documentation and support, but also on other attributes such as modularity, flexibility, reusability, expandability and interoperability.

*Integrated teams of specialists are needed within HEP -* The development and application of accelerator algorithms and codes have become extremely complex and specialized endeavors. The depth and breadth of computational accelerator science and technology (CAST) are both deepening and widening, calling for the development of teams of specialists including computational physicists, applied mathematicians and computer scientists. Such an approach is being adopted elsewhere, e.g. in the Computational Physics Division at the Max-Planck-Institut für Plasmaphysik in Garching Germany, on magnetic fusion [4], and in the project CILEX (or "Interdisciplinary Center on EXtreme Light") in Saclay, France, on laser applications [5]. In those programs, the development of algorithms and codes is assigned to a team of computational physicists, applied mathematicians and computer scientists who cooperate full time on a unique application. Similar teams of dedicated specialists are needed to efficiently respond to the increasing requirements for CAST, to adapt and/or develop algorithms and codes for emerging computer architectures, and to develop codes with high levels of modularity and interoperability.

**Increasing coordination of code development**

*Need for coordination of codes into a cohesive tool set -* Numerous simulation codes have been developed at institutions worldwide for the modeling of particle beams and the design of particle accelerators; while a wide array of options is available to modelers, the tools lack cohesion and there has historically been significant duplication and inefficiency. With a few notable exceptions, the development paradigm for beam physics and accelerator codes has largely been: a code linked to a project or narrow topic, developed by one researcher (usually a physicist), with occasional (subcontracted or collaborative) help from computer scientists and/or applied mathematicians. By its nature, this model has inevitably led to the development of a multitude of highly specialized codes with limited scopes that lack the depth and breadth that are required for integrated start-to-end modeling of complex multi-physics systems. This situation calls for a higher level of coordination among the various modeling and code development efforts, with progressive integration of the various codes into a cohesive tool set.

***Portfolio should cover the range from low- to high-end computing -*** Supercomputers are now most commonly an aggregate of a large number of processing units that are very similar to those found in desktop or laptop personal computers. Thus, it is increasingly the case that even for codes that are ultimately used for production on very large supercomputers, development is done primarily on small clusters, desktops, and even laptops, for faster turnaround and more efficient development, testing and debugging. As a result, a modern accelerator computer code is often used on a wide range of computers ranging from laptops to supercomputers, to solve problems ranging very small to very large in scale. Hence, the separation between algorithms and codes targeted at personal computers versus supercomputers – or serial versus massively parallel architectures – is diminishing and it is essential to envision a comprehensive program of computational accelerator science tools that function well on a broad range of platforms as needed. Heterogeneous computing architecture is becoming ubiquitous on the full range of platforms, presenting great opportunities and challenges in the near term for scientific software.

***Need and solution for non-disruptive integration of codes into common tool set -*** The HEP community has made very significant investments into the existing pool of accelerator modeling tools. It is thus desirable that the development of an integrated solution brings with it as little disruption as possible, while enabling the various components to interact with each other and be expandable. In the past decade, the Python [6,7] scripting language has emerged as a high-level solution for rapid development and prototyping of scientific applications, on par with other high-level tools such as Maple, Matlab or Octave. In addition, Python is expandable and can be easily coupled to the high-performance programming languages, especially FORTRAN, C and C++, which are employed by the vast majority of accelerator codes. A high-level scripting interface for rapid development and prototyping, offering easy interfacing with high performance languages and expandability, represents an ideal solution to the challenge of coupling of existing codes while minimizing disruption and enabling both interoperability and expansion capabilities. With such a construct, existing codes continue unmodified, but their functionalities are exposed to each other and to users. Each code represents a module in a much larger pool that users can access to create multi-physics applications that go well beyond the scope of any individual code or small set of codes. A well-designed, extensible API will be essential for long-term interoperability of these independently developed components. This approach promises to offer the HEP community an integrated tool set of unprecedented power and versatility.

**Summary**

Having identified challenges for maximizing the impact of computer modeling on the design of future particle accelerators and the development of new accelerator techniques & technologies, we have proposed a strategy to meet those challenges. These ideas are broadly consistent with the strategy of the National Science Foundation [8]. This white paper was originally presented at DOE's Community Summer Study, aka "Snowmass on the Mississippi" [9].